# MMAPs to model complex multi-state systems with vacation policies in the repair facility

Juan Eloy Ruiz-Castro[1] and Christian Acal[2]

Department of Statistics and Operations Research and Mathematics Institute IMAG. University of Granada.

[1]e-mail: jeloy@ugr.es  ;  [2]e-mail: chracal@ugr.es

## ABSTRACT

Two complex multi-state systems subject to multiple events are built in an algorithmic and computational way by considering phase-type distributions and Markovian arrival processes with marked arrivals. The internal performance of the system is composed of different degradation levels and internal repairable and non-repairable failures can occur. Also, the system is subject to external shocks that may provoke repairable or non-repairable failure. A multiple vacation policy is introduced in the system for the repairperson. Preventive maintenance is included in the system to improve the behaviour. Two types of task may be performed by the repairperson; corrective repair and preventive maintenance. The systems are modelled, the transient and stationary distributions are built and different performance measures are calculated in a matrix-algorithmic form. Cost and rewards are included in the model in a vector matrix way. Several economic measures are worked out and the net reward per unit of time is used to optimize the system. A numerical example shows that the system can be optimized according to the existence of preventive maintenance and the distribution of vacation time. The results have been implemented computationally with Matlab and R (packages: *expm*, *optim*).

**Keywords.** Phase-type distribution (PH)**;** Marked Markovian arrival process (MMAP); Vacation policy; Preventive maintenance.

## 1. Introduction

The occurrence of repairable and non-repairable failures in a reliability system can provoke severe damage and major financial costs. To avoid such an outcome, several methodologies are considered, such as redundant systems and preventive maintenance.

Preventive maintenance (PM) is the maintenance methodology of systems in order to keep them running and prevent any costly unplanned downtime. A successful maintenance policy requires planning and scheduling maintenance of system before a failure takes place. In this respect, several preventive maintenance policies have been proposed in the reliability field. Barlow and Hunter (1960) considered two types of preventive maintenance policies to optimize a system depending on the failure distribution. Multiple preventive maintenance policies were given in detail in Nakagawa (1977, 2005) and Finkelstein et al.(2020) developed a new model for the hybrid preventive maintenance of systems with partially observable degradation. Recently, other strategies to optimize a reliability system are given in Shi et al. (2022). In this work an advanced estimation strategy is proposed, in which only one surrogate model is built, being able to estimate the failure probabilities of different performance functions.

Nowadays binary systems have been extended by multi-state systems (MSS). Complex systems that have a finite number of performance levels and various failure modes, each producing different effects on system performance, are termed multi-state systems (MSS). Murchland (1975) discussed this concept, which has since been developed extensively. One of the main problems when complex multi-state systems are modelled is that intractable expressions appear in the modelling and in the performance functions. This fact makes difficult the algorithmization and interpretation of results. One possible solution is based on two elements, phase-type distributions (PH) and Markovian arrival processes (MAP), which enable to express complex systems in an algorithmic and computational way. PH were introduced by Neuts (1975) and studied in detail in Neuts (1981). PH has been considered in multiple fields such as queuing theory, physics, reliability and survival. In the physics field PH has been considered to model the behaviour of resistive memories RRAM in Acal et al. (2019). They have also been considered in survival to study the evolution of several illnesses such as cancer (Pérez-Ocón et al, 1998; Ruiz-Castro and Zenga, 2020). The modelling with PH in reliability is extensive. A transient analysis of a multi-state system was modelled by using PH in Pérez-Ocón et al. (2006). One of the main properties of PH is that it is dense in the non-negative probability distributions set. Therefore, any non-negative probability distribution can be approximated so much as it is desirable through a PH.

MAP is a counting process in which PH distributions play an important role. This process was given by Neuts (1979) and reviewed by Artalejo et al. (2010) and He (2014). A special case is that of the MAP with marked arrivals (MMAP), which enables us to count different types of arrival. MMAPs are developed in a compressive form in He (2014). This markovian structure, analogously to PH, enables to count event in an algorithmic way. Multiple examples in several fields are proposed in He and Neuts (1998) and it has also been considered in the modelling of reliability discrete systems (Ruiz-Castro, 2018, 2020).

In complex multi-state systems is usual to consider either repairing immediately after repairable failure or immediate replacement when a non-repairable occurs. However, this might not be the case in a real scenario. For example, a failed unit might not be repaired immediately in a small or medium-sized firm that cannot afford to employ a full-time repairperson. The permanent service facility may increase cost, idleness, deterioration in quality. To reduce the wastage of valuable resources like time, money, quality, etc., vacation is a prominent idea for the service facility. Instead of remaining idle during this period, the repairperson may take a 'vacation' and/or use the time to do other work, thus optimising resources and reducing costs. A repairperson is on vacation when absent from the repair facility, whether or not it is empty. The economic implications of this situation should be considered, taking into account that the vacation policy applied might impact both on performance and also on economic rewards/costs.

Multiple vacation policies have been considered in queuing theory and reliability. A comparative study of different vacation policies on the reliability characteristics is presented in Shekhar et al. (2020). A Markovian queuing model with a vacation policy in the repair facility where the vacation period follows negative exponential is developed in Kalyanaraman and Sundaramoorthy (2019).
Multi-state Markov systems with vacation periods have also been considered. In Zhang et al. (2017) a k-out-of-n system with a single repairperson, assuming a phase-type distribution for the vacation time and an exponential distribution for the lifetime of the units is modelled.

In this paper, two complex multi-state unit systems subject to multiple events, such as internal and external repairable and non-repairable failures are modelled. The internal performance of the system is partitioned into several levels of degradation. A vacation policy is introduced by considering the internal degradation levels. The first system is extended to include preventive maintenance and the corresponding vacation policy. The repairperson performs two different tasks, corrective repair and preventive maintenance. The corrective and preventive time distributions can be different for both cases. The system is modelled by using PHs and MMAPs in an algorithmic and computational way. Multiple measures such as availability, reliability function, rate of occurrence of failure (ROCOF) and mean number of events are worked out. The transient and stationary distributions are calculated in a matrix-algorithmic form. Costs and rewards, depending on the internal degradation levels, are included. Everything is algorithmically and computationally modelled and has been applied to compare and optimize two similar systems with and without preventive maintenance. The results have been implemented computationally with Matlab and R-cran.

The paper is organized as follows. Both systems are described in Section 2. These systems are modelled in Section 3. Measures and costs/profits are developed in Sections 4 and 5, respectively. In Section 6 a numerical application is given where optimization and comparison are shown. Conclusions are given in Section 7. Finally two Appendices show the algorithm of the models in detail.

## 2. The systems

Two different multi-state one-unit systems are modelled by considering Markovian Arrival Processes with marked arrivals, with and without preventive maintenance. Both systems are subject to internal degradation and external shocks. Repairable and non-repairable failures, depending on the internal degradation state, can occur in both cases. The systems can be observed only by the repairperson and this one is not always at his workplace. A policy of multiple vacation periods for the repairperson is given.

### 2.1. The system without preventive maintenance

The internal behaviour of the unit consists of two different levels, minor degradation and moderate degradation. The number of states for the first and second level is $n_1$ and $n_2$ respectively. The unit can suffer repairable and non-repairable failures from both degradation levels. The plug in which unit is connected can undergo external shocks that can also provoke repairable or non-repairable failures. The system is composed of one repairperson. This repairperson can take multiple vacation periods depending on the unit degradation level. Thus, the repairperson is on vacation initially. When this one returns, if the system is in minor degradation level then a new random vacation period will be started by the repairperson. Otherwise this subject stays at the workplace waiting for a possible repairable failure. If the repairperson is on vacations and a repairable failure occurs, it remains in repairable failure macro-state till the repairperson returns and then he begins the repair. Analogously, if the repairperson is on vacations and a non-repairable failure occurs then it will be replaced by an identical unit when the repairperson returns. On the contrary, if the repairperson is at the workplace and a repairable or non-repairable occurs, then the repair starts immediately or it is replaced in a negligible time respectively. All random times embedded in the model go through different states until the event occurs.

These times embedded in the system verify the following assumptions.

**Assumption 1.** The internal operational time follows a PH distribution with representation $(\boldsymbol{\alpha}, \mathbf{T})$ with order $n$. The $n$ phases are partitioned into two macro-states, minor degradation (first $n_1$ phases) and moderate degradation level (remaining phases, $n_2$). The PH representation is composed of matrix blocks according to the levels. Then, $\boldsymbol{\alpha}_1$ is a row vector composed of the first $n_1$ elements of $\boldsymbol{\alpha}$. The matrix $\mathbf{T}$ is given by

$$\mathbf{T} = \left( \begin{array}{c|c} \mathbf{T}_{11} & \mathbf{T}_{12} \\ \hline \mathbf{0} & \mathbf{T}_{22} \end{array} \right).$$

The order of $\mathbf{T}_{11}$ and $\mathbf{T}_{22}$ is $n_1$ and $n_2$ respectively.

The column vector $\mathbf{T}^0 = -\mathbf{T}\mathbf{e}$ contains the failure rates from the different operational phases. Throughout this paper $\mathbf{e}$ is a column vector of ones with appropriate order and $\mathbf{A}^0 = -\mathbf{A}\mathbf{e}$ for any matrix $\mathbf{A}$.

The column vector is expressed as $\mathbf{T}^0 = \mathbf{T}^0_r + \mathbf{T}^0_{nr}$ where $\mathbf{T}^0_r$ and $\mathbf{T}^0_{nr}$ are column vectors which contain the repairable and non-repairable failure rates from the operational phases, respectively. These vectors are partitioned according to the degradation levels as $\mathbf{T}^0_{i,r}$ and $\mathbf{T}^0_{i,nr}$ for $i = 1, 2$.

**Assumption 2.** The external shock is modelled through a PH renewal process where the time between two consecutive shocks is PH distributed with representation $(\boldsymbol{\gamma}, \mathbf{L})$. The order of this representation is $p$. The vector $\mathbf{L}^0$ contains the transition intensities up to external shock rate depending on the phases of external shock time. This vector is partitioned as $\mathbf{L}^0 = \mathbf{L}^0_r + \mathbf{L}^0_{nr}$ where $\mathbf{L}^0_r$ and $\mathbf{L}^0_{nr}$ are column vectors which contain the repairable and non-repairable external shock rates, respectively.

**Assumption 3.** The vacation time follows a PH distribution with representation $(\boldsymbol{\upsilon}, \mathbf{V})$, being $\mathbf{V}$ a matrix of order $v$.

**Assumption 4.** The correction repair time follows a PH distribution with representation $(\boldsymbol{\beta}^1, \mathbf{S}_1)$, with $\mathbf{S}_1$ being a matrix of order $m_1$.

Therefore, the behaviour of the system can be partitioned into six macro-states of the state-space $S$,

$$S = \{E_1 = O_1, E_2 = O_2^{WR}, E_3 = O_2^R, E_4 = RF^{WR}, E_5 = NRF^{WR}, E_6 = CR\}.$$

These macro-states contain the phases with the following situations:

- $E_1 = O_1$: The unit is working in minor internal degradation.

$$E_1 = O_1 = \{(i, j, k); i = 1, \ldots, n_1, j = 1, \ldots, p, k = 1, \ldots, v\}$$

  $i$: phase of the minor internal degradation level

  $j$: phase of the external shock time

  $k$: phase of the vacation time

- $E_2 = O_2^{WR}$: The unit is working in middle internal degradation with the repairperson on vacation. The superscript *WR* indicates "without repairperson" in the repair facility,

$$E_2 = O_2^{WR} = \{(i, j, k); i = 1, \ldots, n_2, j = 1, \ldots, p, k = 1, \ldots, v\}$$

  $i$: phase of the middle internal degradation level

  $j$: phase of the external shock time

  $k$: phase of the vacation time

- $E_3 = O_2^R$: The unit is working in middle internal degradation with the repairperson on the workplace. The superscript *R* indicates that the "repairperson" is in the repair facility,

$$E_3 = O_2^R = \{(i, j); i = 1, \ldots, n_2, j = 1, \ldots, p\}$$

  $i$: phase of the middle internal degradation level

  $j$: phase of the external shock time

- $E_4 = RF^{WR}$: The unit is broken with repairable failure and the repairperson is on vacation. *RF* indicates that the system is in "repairable failure" and the superscript indicates "without repairperson" in the repair facility,

$$E_4 = RF^{WR} = \{(j, k); j = 1, \ldots, p, k = 1, \ldots, v\}$$

  $j$: phase of the external shock time

  $k$: phase of the vacation time

- $E_5 = NRF^{WR}$: The unit is broken with non-repairable failure and the repairperson is on vacation. *NRF* indicates that the unit is in "non-repairable failure" and the superscript indicates "without repairperson" in the repair facility

$$E_5 = NRF^{WR} = \{(j,k); j = 1,\ldots, p; k = 1,\ldots, v\}$$

  *j*: phase of the external shock time
  *k*: phase of the vacation time

- $E_6 = CR$: The unit is on "corrective repair" with the repairperson

$$E_6 = CR = \{(j,l); j = 1,\ldots, p; l = 1,\ldots, m_1\}$$

  *j*: phase of the external shock time
  *l*: phase of the corrective repair

The system will be operational while it occupies a state of the macro-state $W = \{E_1 = O_1, E_2 = O_2^{WR}, E_3 = O_2^{R}\}$ and it will be non-operational when it is found in some macro-state of $F = \{E_4 = RF^{WR}, E_5 = NRF^{WR}, E_6 = CR\}$.

## 2.2. The system with preventive maintenance

The system described in section above is extending by including preventive maintenance. In this case we assume that the internal behaviour is composed of three different levels, i.e., minor, middle and major degradation. The number of states is $n_1$, $n_2$ and $n_3$ for these levels respectively. External shocks with similar consequences are also included in this model. The vacation time policy is different for this system with preventive maintenance. The repairperson is also on vacation initially. When this one returns, the repairperson can observe five different situations instead of four.

- Minor internal degradation level. The repairperson begins a new random vacation period.
- Middle internal degradation. The repairperson stays at the workplace waiting for a possible repairable failure.
- Major internal damage. The repairperson starts the preventive maintenance.
- Repairable failure. The repairperson begins the corrective repair.
- Non-repairable failure. The repairperson replaces the unit by a new and identical one in a negligible time.

Other possibility is that a repairable failure or a non-repairable failure occurs while the repairperson is at the workplace without working. In this case, the corrective repair begins or the unit is replaced immediately after occurring the event, respectively.

When preventive maintenance is considered, the times embedded in the system verify the following assumptions.

**Assumption 1.** The internal operational time follows a PH distribution with representation $(\boldsymbol{\alpha}, \mathbf{T})$ with order $n$. The $n$ phases are partitioned into three macro-states, minor degradation level (first $n_1$ phases), middle degradation level (the following first $n_2$ phases) and major degradation level (last $n_3$ phases). The PH representation is composed of matrix blocks according to the levels. Then, $\boldsymbol{\alpha}_1$ is a row vector composed of the first $n_1$ elements of $\boldsymbol{\alpha}$. The matrix $\mathbf{T}$ is given by

$$\mathbf{T} = \begin{pmatrix} \mathbf{T}_{11} & \mathbf{T}_{12} & \mathbf{T}_{13} \\ \mathbf{0} & \mathbf{T}_{22} & \mathbf{T}_{23} \\ \mathbf{0} & \mathbf{0} & \mathbf{T}_{11} \end{pmatrix}.$$

The order of $\mathbf{T}_{11}$, $\mathbf{T}_{22}$ and $\mathbf{T}_{33}$ is $n_1$, $n_2$ and $n_3$, respectively.

The column vector $\mathbf{T}^0$ is expressed as $\mathbf{T}^0 = \mathbf{T}_r^0 + \mathbf{T}_{nr}^0$ again. These vectors are partitioned according to the degradation levels as $\mathbf{T}_{i,r}^0$ and $\mathbf{T}_{i,nr}^0$ for $i = 1,2,3$.

**Assumption 2.** The same assumption 2 that the one given for the system without preventive maintenance.

**Assumption 3.** The same assumption 3 that the one given for the system without preventive maintenance.

**Assumption 4.** The same assumption 4 that the one given for the system without preventive maintenance.

**Assumption 5.** The preventive maintenance time follows a PH distribution with representation $(\boldsymbol{\beta}^2, \mathbf{S}_2)$, with $\mathbf{S}_2$ being a matrix of order $m_2$.

When preventive maintenance is included, eight macro-states are possible. The macro-state space for this layout is the following,

$$S = \{E_1 = O_1, E_2 = O_2^{WR}, E_3 = O_2^{R}, E_4 = O_3^{WR}, E_5 = RF^{WR}, E_6 = NRF^{WR}, E_7 = PM, E_8 = CR\}$$
.

The macro-states $E_1$, $E_2$ and $E_3$ are the same as for the case without preventive maintenance. The macro-states $E_5$, $E_6$ and $E_8$ defined in the current system are the macro-states $E_4$, $E_5$ and $E_6$, respectively. Then the new macro-states for this system are $E_4 = O_3^{WR}$, and $E_7 = PM$.

These new macro-states contain the phases with the following situations:

- $E_4 = O_3^{WR}$: The unit is working in major internal degradation. The superscript indicates "without repairperson" in the repair facility,

$$E_4 = O_3^{WR} = \{(i, j, k); i = 1, \ldots, n_3, j = 1, \ldots, p, k = 1, \ldots, v\}$$

  $i$: phase of the major internal degradation level

  $j$: phase of the external shock time

  $k$: phase of the vacation time

- $E_7 = PM$: The unit is in "preventive maintenance".

$$E_7 = PM = \{(j, l); j = 1, \ldots, p; l = 1, \ldots, m_2\}$$

  $j$: phase of the external shock time

  $l$: phase of the preventive maintenance time

For the system with preventive maintenance, the operational macro-state is $W = \{E_1 = O_1, E_2 = O_2^{WR}, E_3 = O_2^{R}, E_4 = O_3^{WR}\}$ and the non-operational is given by $F = \{E_5 = RF^{WR}, E_6 = NRF^{WR}, E_7 = PM, E_8 = CR\}$.

## 3. Modelling the systems through Marked Markovian Arrival Processes

The systems described in Section 2 are modelled through Markovian Arrival Processes with marked arrivals. These models enable us not only to analyse the system evolution, but also the number of different events can be worked out over time. The model for the system with preventive maintenance is developed in this work. The case for without preventive maintenance is given in Appendix A.

The multi-state unit may undergo the following types of events which are denoted as,

*O*: No events (no PM, no failure, no end of vacation)

*RF+CR*: Repairable failure and start of corrective repair (the repairperson was at the workplace)

*RF*: Only repairable failure (the repairperson continues on vacations)

*PM*: preventive maintenance (the repairperson continues on vacations)

*NRF+NU*: Non-repairable failure and immediate replacement (immediate new unit because the repairperson was at the workplace)

*NRF*: Only non-repairable failure (the repairperson continues on vacations)

*I*: Only return from vacations

*I+PM*: Return from vacations and start preventive maintenance (the unit is at major degradation level)

*I+CR*: Return from vacations and start corrective repair (the unit was in *RF*)

*I+NU*: Return from vacations and immediate replacement (immediate new unit because the repairperson was at the workplace)

## 3.1. The MMAP

The MMAP associated to the system has been built according to the different events aforementioned. The representation is given by

$$\left(\mathbf{D}^0, \mathbf{D}^{RF+CR}, \mathbf{D}^{RF}, \mathbf{D}^{NRF+NU}, \mathbf{D}^{NRF}, \mathbf{D}^{I}, \mathbf{D}^{I+CR}, \mathbf{D}^{I+NU}\right),$$

where $\mathbf{D}^Y$ contains the transition intensities for the event *Y*. These matrices are composed of matrix blocks. Each matrix block contains the transition intensities for the event *Y* by considering the macro-states of the state-space *S*.

The block matrices for the events *RF* and *RF+CR* are described next. The remainder are given in Appendix B.

**Matrix block $\mathbf{D}^{RF}$**

The matrix block $\mathbf{D}^{RF}$ contains the transitions intensities from an operational state to a repairable failure (without other event). Therefore, it is only possible for the transitions between the macro-states $O_1 \to RF$, $O_2^{WR} \to RF$ or $O_3^{WR} \to RF$. The block $\mathbf{D}^{RF}$ is given by,

$$\mathbf{D}^{RF} = \begin{pmatrix} 0 & 0 & 0 & 0 & \mathbf{C}_{O_1 RF} & 0 & 0 & 0 \\ 0 & 0 & 0 & 0 & \mathbf{C}_{O_2^{WR} RF} & 0 & 0 & 0 \\ 0 & 0 & 0 & 0 & 0 & 0 & 0 & 0 \\ 0 & 0 & 0 & 0 & \mathbf{C}_{O_3^{WR} RF} & 0 & 0 & 0 \\ 0 & 0 & 0 & 0 & 0 & 0 & 0 & 0 \\ 0 & 0 & 0 & 0 & 0 & 0 & 0 & 0 \\ 0 & 0 & 0 & 0 & 0 & 0 & 0 & 0 \\ 0 & 0 & 0 & 0 & 0 & 0 & 0 & 0 \end{pmatrix}.$$

The matrix $\mathbf{C}_{O_1 RF}$ contains the transition between the macro-states $O_1 \to RF$. It occurs when an internal repairable failure takes place and the external shock and the vacation times do not change ($\mathbf{T}_{1,r}^0 \otimes \mathbf{I} \otimes \mathbf{I}$), or because an external shock occurs by provoking a repairable failure ($\mathbf{e} \otimes \mathbf{L}_r^0 \boldsymbol{\gamma} \otimes \mathbf{I}$). In the last case, the internal damage finishes and the vacation time is not altered. Then,

$$\mathbf{C}_{O_1 RF} = \mathbf{T}_{1,r}^0 \otimes \mathbf{I} \otimes \mathbf{I} + \mathbf{e} \otimes \mathbf{L}_r^0 \boldsymbol{\gamma} \otimes \mathbf{I}.$$

The matrix $\mathbf{C}_{O_2^{WR} RF}$ contains the transition between the macro-states $O_2^{WR} \to RF$. It occurs when an internal repairable failure takes place from middle degradation level without repair and the external shock and the vacation time remain identical ($\mathbf{T}_{2,r}^0 \otimes \mathbf{I} \otimes \mathbf{I}$), or because an external shock occurs by provoking a repairable failure ($\mathbf{e} \otimes \mathbf{L}_r^0 \boldsymbol{\gamma} \otimes \mathbf{I}$). Then,

$$\mathbf{C}_{O_2^{WR} RF^{WR}} = \mathbf{T}_{2,r}^0 \otimes \mathbf{I} \otimes \mathbf{I} + \mathbf{e} \oplus \mathbf{L}_r^0 \boldsymbol{\gamma} \otimes \mathbf{I}.$$

Finally, The matrix $\mathbf{C}_{O_3^{WR} RF}$ contains the transition between the macro-states $O_3^{WR} \to RF$. The reasoning is similar as for the case above but from the macro-state in major degradation level without repair. It is given by,

$$\mathbf{C}_{O_3^{WR} RF^{WR}} = \mathbf{T}_{3,r}^0 \otimes \mathbf{I} \otimes \mathbf{I} + \mathbf{e} \otimes \mathbf{L}_r^0 \boldsymbol{\gamma} \otimes \mathbf{I}.$$

**Matrix block $\mathbf{D}^{RF+CR}$**

The matrix block $\mathbf{D}^{RF+CR}$ contains the transitions intensities from an operational state to repairable failure (with immediate corrective repair). Therefore, it is only possible for the transitions between the macro-states $O_2^R \to RF$ because the repairperson must be at the workplace. This matrix block is

$$\mathbf{D}^{RF+CR} = \begin{pmatrix} 0 & 0 & 0 & 0 & 0 & 0 & 0 & 0 \\ 0 & 0 & 0 & 0 & 0 & 0 & 0 & 0 \\ 0 & 0 & 0 & 0 & 0 & 0 & 0 & \mathbf{C}_{O_2^R CR} \\ 0 & 0 & 0 & 0 & 0 & 0 & 0 & 0 \\ 0 & 0 & 0 & 0 & 0 & 0 & 0 & 0 \\ 0 & 0 & 0 & 0 & 0 & 0 & 0 & 0 \\ 0 & 0 & 0 & 0 & 0 & 0 & 0 & 0 \\ 0 & 0 & 0 & 0 & 0 & 0 & 0 & 0 \end{pmatrix}.$$

The matrix $\mathbf{C}_{O_2^R RF}$ contains the transition between the macro-states $O_2^R \to RF$. The repairperson is at the workplace and a repairable failure occurs (internal or external) from middle degradation level. After the repairable repair occurs, a corrective repair begins given that the repairperson is prepared for that. The remainder does not change. Then,

$$\mathbf{C}_{O_2^R CR} = \mathbf{T}_{2,r}^0 \otimes \mathbf{I} \otimes \boldsymbol{\beta}^1 + \mathbf{e} \otimes \mathbf{L}_r^0 \boldsymbol{\gamma} \otimes \boldsymbol{\beta}^1.$$

## 3.2. Transient distribution

The system is modelled by the MMAP given in Section 3.1. Therefore, the Q-matrix associated with the Markov process by which the system is governed adopts the expression

$$\mathbf{D} = \mathbf{D}^0 + \mathbf{D}^{RF+CR} + \mathbf{D}^{RF} + \mathbf{D}^{NRF+NU} + \mathbf{D}^{NRF} + \mathbf{D}^I + \mathbf{D}^{I+CR} + \mathbf{D}^{I+NU}.$$

We assume that the system is new and the repairperson in on vacation initially. Therefore, the initial distribution for the system with preventive maintenance is given by $\boldsymbol{\theta} = (\boldsymbol{\alpha} \otimes \boldsymbol{\omega} \otimes \boldsymbol{\upsilon}, \mathbf{0})$ respectively, with $\boldsymbol{\omega}$ being the stationary distribution of the external failure. This fact is assumed because external shocks happen in a continuous way. Therefore, $\boldsymbol{\omega} = (\mathbf{0}, 1)\left(\left(\mathbf{L} + \mathbf{L}^0 \boldsymbol{\gamma}\right)^* | \mathbf{e}\right)^{-1}$.

The transient distribution probability is worked out from $\mathbf{P}(t) = \exp(\mathbf{D}t)$ and the probability of being at any phase of the macro-state $E_i$, that is, $\mathbf{p}_{E_i}(t)$, is given by the vector $\mathbf{p}(t) = \boldsymbol{\theta} \exp(\mathbf{D}t)$ restricted to the elements of the macro-state $E_i$.

## 3.3. The stationary distribution

The stationary distribution is denoted as $\boldsymbol{\pi}$ and it is partitioned according to the macro-state space $S$. Therefore, it is denoted as $\boldsymbol{\pi}_i$ to the vector $\boldsymbol{\pi}_i = \lim_{t \to \infty} \mathbf{p}_{E_i}(t)$. To ease the development, the generator of the process is denoted as

$$\mathbf{D} = \begin{pmatrix} \mathbf{D}_{11} & \mathbf{D}_{12} & \mathbf{0} & \mathbf{D}_{14} & \mathbf{D}_{15} & \mathbf{D}_{16} & \mathbf{0} & \mathbf{0} \\ \mathbf{0} & \mathbf{D}_{22} & \mathbf{D}_{23} & \mathbf{D}_{24} & \mathbf{D}_{25} & \mathbf{D}_{26} & \mathbf{0} & \mathbf{0} \\ \mathbf{D}_{31} & \mathbf{0} & \mathbf{D}_{33} & \mathbf{0} & \mathbf{0} & \mathbf{0} & \mathbf{D}_{37} & \mathbf{D}_{38} \\ \mathbf{0} & \mathbf{0} & \mathbf{0} & \mathbf{D}_{44} & \mathbf{D}_{45} & \mathbf{D}_{46} & \mathbf{D}_{47} & \mathbf{0} \\ \mathbf{0} & \mathbf{0} & \mathbf{0} & \mathbf{0} & \mathbf{D}_{55} & \mathbf{0} & \mathbf{0} & \mathbf{D}_{58} \\ \mathbf{D}_{61} & \mathbf{0} & \mathbf{0} & \mathbf{0} & \mathbf{0} & \mathbf{D}_{66} & \mathbf{0} & \mathbf{0} \\ \mathbf{D}_{71} & \mathbf{0} & \mathbf{0} & \mathbf{0} & \mathbf{0} & \mathbf{0} & \mathbf{D}_{77} & \mathbf{0} \\ \mathbf{D}_{81} & \mathbf{0} & \mathbf{0} & \mathbf{0} & \mathbf{0} & \mathbf{0} & \mathbf{0} & \mathbf{D}_{88} \end{pmatrix}.$$

As it is well known, the stationary distribution is the solution of the matrix balance equation $\boldsymbol{\pi} \cdot \mathbf{D} = \mathbf{0}$ with the normalization condition $\boldsymbol{\pi} \cdot \mathbf{e} = 1$.

In a matrix way, the balance equations are given by

$$\pi_1 D_{11} + \pi_3 D_{31} + \pi_6 D_{61} + \pi_7 D_{71} + \pi_8 D_{81} = 0$$

$$\pi_1 D_{12} + \pi_2 D_{22} = 0$$

$$\pi_2 D_{23} + \pi_3 D_{33} = 0$$

$$\pi_1 D_{14} + \pi_2 D_{24} + \pi_4 D_{44} = 0$$

$$\pi_1 D_{15} + \pi_2 D_{25} + \pi_4 D_{45} + \pi_5 D_{55} = 0$$

$$\pi_1 D_{16} + \pi_2 D_{26} + \pi_4 D_{41} + \pi_6 D_{66} = 0$$

$$\pi_3 D_{37} + \pi_4 D_{47} + \pi_7 D_{77} = 0$$

$$\pi_3 D_{38} + \pi_5 D_{58} + \pi_8 D_{88} = 0$$

$$\pi \cdot e = 1$$

The solution of this matrix system is given by

$$\pi_i = \pi_1 R_i \quad ; \quad i = 2,\ldots,8$$

with

$$R_2 = G_{12}$$

$$R_3 = R_2 G_{23}$$

$$R_4 = G_{14} + R_2 G_{24}$$

$$R_5 = G_{15} + R_2 G_{25} + R_4 G_{45}$$

$$R_6 = G_{16} + R_2 G_{26} + R_4 G_{46}$$

$$R_7 = R_3 G_{37} + R_4 G_{47}$$

$$R_8 = R_3 G_{38} + R_5 G_{58}.$$

Being $G_{jk} = -D_{jk} D_{kk}^{-1}$ for the corresponding case.

The $\pi_1$ is achieved from the first and last matrix equation,

$$\pi_1 D_{11} + \pi_3 D_{31} + \pi_6 D_{61} + \pi_7 D_{71} + \pi_8 D_{81} = 0$$
$$\pi \cdot e = 1$$

Then,

$$\pi_1 = (0,1)\left( A^* \mid \left( I + \sum_{a=2}^{8} R_a \right) e \right)^{-1},$$

with $A^*$ being the matrix $A$ without the first column and $A = D_{11} + R_3 D_{31} + R_6 D_{61} + R_7 D_{71} + R_8 D_{81}$.

## 4. Measures

Several interesting measures in the reliability field such as ROCOF, availability, reliability and several mean number of events are worked out in this section.

### 4.1. Availability

The availability is the probability of being operational the system at a certain time. It is given by

$$A(t) = \sum_{i=1}^{4} \mathbf{p}_{E_i}(t)\mathbf{e}.$$

For the stationary case it is $A = \sum_{i=1}^{4} \boldsymbol{\pi}_i \mathbf{e}$.

### 4.2. Reliability

Several reliability functions may be defined for this system (time up to repairable failure, time up to non-repairable failure or time up to first case that the system in not operational). We define it as the first time that the unit is not operational. The probability distribution of this time is PH with representation $(\boldsymbol{\theta}', \mathbf{D}')$ being

$$\boldsymbol{\theta}' = (\boldsymbol{\alpha} \otimes \boldsymbol{\omega} \otimes \boldsymbol{\upsilon}, \mathbf{0}) \; ; \; \mathbf{D}' = \begin{pmatrix} \mathbf{C}^1_{O_1 O_1} + \mathbf{C}^2_{O_1 O_1} & \mathbf{C}_{O_1 O_2^{WR}} & \mathbf{0} & \mathbf{C}_{O_1 O_3^{WR}} \\ \mathbf{0} & \mathbf{C}_{O_2^{WR} O_2^{WR}} & \mathbf{C}_{O_2^{WR} O_2^{R}} & \mathbf{C}_{O_2^{WR} O_3^{WR}} \\ \mathbf{0} & \mathbf{0} & \mathbf{C}_{O_2^{R} O_2^{R}} & \mathbf{0} \\ \mathbf{0} & \mathbf{0} & \mathbf{0} & \mathbf{C}_{O_3^{WR} O_3^{WR}} \end{pmatrix}.$$

The reliability function is given by $R(t) = \boldsymbol{\theta}' \exp(\mathbf{D}'t) \cdot \mathbf{e}$.

### 4.3. ROCOF$_{RF}$ and ROCOF$_{NRF}$ (rate of occurrence of repairable and non-repairable failure)

The rate of occurrence of repairable failure is the rate of undergoing a repairable failure at a certain time $t$. It is given by

$$ROCOF_{RF}(t) = \mathbf{p}_{E_1}(t) \cdot \mathbf{C}_{E_1 E_5} \cdot \mathbf{e} + \mathbf{p}_{E_2}(t) \cdot \mathbf{C}_{E_2 E_5} \cdot \mathbf{e} + \mathbf{p}_{E_3}(t) \cdot \mathbf{C}_{E_3 E_8} \cdot \mathbf{e} + \mathbf{p}_{E_4}(t) \cdot \mathbf{C}_{E_4 E_5} \cdot \mathbf{e}.$$

In stationary regime it is

$$ROCOF_{RF} = \boldsymbol{\pi}_1 \cdot \mathbf{C}_{E_1 E_5} \cdot \mathbf{e} + \boldsymbol{\pi}_2 \cdot \mathbf{C}_{E_2 E_5} \cdot \mathbf{e} + \boldsymbol{\pi}_3 \cdot \mathbf{C}_{E_3 E_8} \cdot \mathbf{e} + \boldsymbol{\pi}_4 \cdot \mathbf{C}_{E_4 E_5} \cdot \mathbf{e}.$$

Analogously for the non-repairable case the rate of occurrence of non-repairable failure is defined as the rate of undergoing a non-repairable failure at a certain time $t$. It is given by

$$ROCOF_{NRF}(t) = \mathbf{p}_{E_1}(t) \cdot \mathbf{C}_{E_1 E_6} \cdot \mathbf{e} + \mathbf{p}_{E_2}(t) \cdot \mathbf{C}_{E_2 E_6} \cdot \mathbf{e} + \mathbf{p}_{E_3}(t) \cdot \mathbf{C}_{E_3 E_1} \cdot \mathbf{e} + \mathbf{p}_{E_4}(t) \cdot \mathbf{C}_{E_4 E_6} \cdot \mathbf{e}$$

This measure in steady-state is

$$ROCOF_{NRF}(t) = \boldsymbol{\pi}_1 \cdot \mathbf{C}_{E_1 E_6} \cdot \mathbf{e} + \boldsymbol{\pi}_2 \mathbf{C}_{E_2 E_6} \cdot \mathbf{e} + \boldsymbol{\pi}_3 \cdot \mathbf{C}_{E_3 E_1} \cdot \mathbf{e} + \boldsymbol{\pi}_4 \cdot \mathbf{C}_{E_4 E_6} \cdot \mathbf{e}$$

### 4.4. Mean number of events

The mean number of events described in Section 3 is worked out from the MMAP. Given an event $Y$, the mean number of events up to time $t$ is given by

$$MN^Y(t) = \boldsymbol{\theta} \int_0^t \exp(\mathbf{D}t) dt \mathbf{D}^Y \mathbf{e} = \boldsymbol{\theta} \left( \exp(\mathbf{D}t) - \mathbf{I} - t\mathbf{e}\boldsymbol{\pi} \right) (\mathbf{D} - \mathbf{e}\boldsymbol{\pi})^{-1} \mathbf{D}^Y \mathbf{e}.$$

From this expression the mean number of even per unit of time in stationary regime is

$$MN^Y = \lim_{t \to \infty} \frac{MN^Y(t)}{t} = \lim_{t \to \infty} \frac{\boldsymbol{\theta} \int_0^t \exp(\mathbf{D}t) dt \mathbf{D}^Y \mathbf{e}}{t} = \boldsymbol{\pi} \mathbf{D}^Y \mathbf{e}.$$

Therefore, depending on $\mathbf{D}^A$, the following measures are calculated.

*Mean number of repairable failures*: $\mathbf{D}^Y = \mathbf{D}^{RF+CR} + \mathbf{D}^{RF}$

*Mean number of non-repairable failures*: $\mathbf{D}^Y = \mathbf{D}^{NRF} + \mathbf{D}^{NRF+NU}$

*Mean number of preventive maintenance*: $\mathbf{D}^Y = \mathbf{D}^{PM} + \mathbf{D}^{I+PM}$

*Mean number of corrective repairs*: $\mathbf{D}^Y = \mathbf{D}^{RF+CR} + \mathbf{D}^{I+CR}$

*Mean number of incorporations*: $\mathbf{D}^Y = \mathbf{D}^{I+PM} + \mathbf{D}^{I+NU} + \mathbf{D}^I + \mathbf{D}^{I+CR}$

*Mean number of new units*: $\mathbf{D}^Y = \mathbf{D}^{I+NU} + \mathbf{D}^{NRF+NU}$

### 5. Cost and rewards

The system described is subject to different events that can provoke costs and rewards according to the macro-states defined. Each time that the system is operational, a reward equal to $B$ is achieved, and analogously, each time that the system is not operational a cost equal to $A$ is produced. Also, a cost is produced while the system is operational. This cost depends on the phases of the internal degradation level. It is given by the column vectors $\mathbf{c}_1$, $\mathbf{c}_2$ and $\mathbf{c}_3$ for minor, middle and major level, respectively.

If the system is in macro-state *PM* or *CR*, the repairperson produces a cost per unit of time depending on the corresponding repairing phases. This cost is given by the vectors $\mathbf{r}_{PM}$ and $\mathbf{r}_{CR}$ respectively. Also, if the repairperson is at the workplace, but idles, a cost equal to $\mathbf{r}_S$ per unit of time is produced.

The net reward vectors according to the macro-states are given as follows,

$$\mathbf{nr}_{O_1} = B\mathbf{e}_{n_1 \cdot p \cdot v} - \mathbf{c}_1 \otimes \mathbf{e}_{p \cdot v}, \quad \mathbf{nr}_{O_2^{WR}} = B\mathbf{e}_{n_2 \cdot p \cdot v} - \mathbf{c}_2 \otimes \mathbf{e}_{p \cdot v}, \quad \mathbf{nr}_{O_2^R} = (B - r_S)\mathbf{e}_{n_2 \cdot p} - \mathbf{c}_2 \otimes \mathbf{e}_p,$$

$$\mathbf{nr}_{O_3^{WR}} = B\mathbf{e}_{n_3 \cdot p \cdot v} - \mathbf{c}_3 \otimes \mathbf{e}_{p \cdot v}, \quad \mathbf{nr}_{RF} = -A\mathbf{e}_{p \cdot v}, \quad \mathbf{nr}_{NRF} = -A\mathbf{e}_{p \cdot v}, \quad \mathbf{nr}_{PM} = -A\mathbf{e}_{p \cdot m_2} - \mathbf{e}_p \otimes \mathbf{c}_{PM}$$

$$\mathbf{nr}_{CR} = -A\mathbf{e}_{p \cdot m_1} - \mathbf{e}_p \otimes \mathbf{c}_{CR}.$$

The net reward vector by considering the phases of the system is,

$$\mathbf{nr} = \begin{pmatrix} \mathbf{nr}_{O_1} \\ \mathbf{nr}_{O_2^{WR}} \\ \mathbf{nr}_{O_2^R} \\ \mathbf{nr}_{O_3^{WR}} \\ \mathbf{nr}_{RF^{WR}} \\ \mathbf{nr}_{NRF^{WR}} \\ \mathbf{nr}_{PM} \\ \mathbf{nr}_{CR} \end{pmatrix}.$$

Defined the net reward vector according the state space, the expected net reward up to a certain time $t$ id given by

$$\Phi(t) = \boldsymbol{\theta} \int_0^t \mathbf{P}(t) dt \cdot \mathbf{nr}.$$

This measure per unit of time is $\dfrac{\Phi(t)}{t}$ and this value in stationary regime is given by $\Phi = \boldsymbol{\pi} \cdot \mathbf{nr}$. It can be interpreted as the net reward per unit of time when the system is balanced.

Other costs associated with different events are added in the model. These are,

$f_{NU}$: fix cost per new unit

$f_{CR}$: fix cost per corrective repair

$f_{PM}$: fix cost per preventive maintenance

$f_I$: fix cost per incorporation from vacation

Therefore, the total expected net reward up to a certain time $t$ is

$$\Psi(t) = \Phi(t) - (1 + MT^{NU}(t)) f_{NU} - MT^{CR}(t) f_{CR} - MT^{PM}(t) f_{PM} - MT^I(t) f_I.$$

This measure per unit of time up to a certain time $t$ is $\Gamma(t) = \dfrac{\Psi(t)}{t}$, and this value in stationary regime is

$$\Gamma = \lim_{t \to \infty} \dfrac{\Psi(t)}{t} = \Phi - MT^{NU} f_{NU} - MT^{CR} f_{CR} - MT^{PM} f_{PM} - MT^{I} f_{I}.$$

## 6. Numerical example: An optimization problem

One interesting problem in reliability is the optimization of systems. In this section, two similar systems, with and without preventive maintenance, are optimised according to the vacation time distribution. Both cases are developed and the optimum systems are compared. The general system consists of multiple internal stages and they are partitioned into minor, middle and major degradation level depending on the damage. In particular, there are seven states partitioned as follows: 1-2, minor degradation level; 3-4, middle degradation level and 5-6-7, major degradation level (if it is observed, the repairperson sends it to preventive maintenance). The repair facility is composed of one sole repairperson. This repairperson can take vacations and the vacation time is random for the general case.

The internal operational time is PH distributed with representation $(\boldsymbol{\alpha}, \mathbf{T})$ where $\boldsymbol{\alpha} = (1,0,0,0,0,0,0)$ and

$$\mathbf{T} = \left( \begin{array}{cc|cc|ccc} -1 & 0.51 & 0.24 & 0.25 & 0 & 0 & 0 \\ 1.2 & -2 & 0.5 & 0.3 & 0 & 0 & 0 \\ \hline 0 & 0 & -0.8 & 0.2 & 0 & 0.16 & 0.16 \\ 0 & 0 & 0.225 & -0.9 & 0.11 & 0.11 & 0.14 \\ \hline 0 & 0 & 0 & 0 & -0.4 & 0.03 & 0.07 \\ 0 & 0 & 0 & 0 & 0.1 & -0.9 & 0.125 \\ 0 & 0 & 0 & 0 & 0.07 & 0.03 & -0.4 \end{array} \right).$$

The internal repairable and non-repairable failure is governed by the following column vectors respectively,

$$\mathbf{T}_r^0 = \begin{pmatrix} 0 \\ 0 \\ \hline 0.24 \\ 0.27 \\ \hline 0.28 \\ 0.63 \\ 0.28 \end{pmatrix} \text{ and } \mathbf{T}_{nr}^0 = \begin{pmatrix} 0 \\ 0 \\ \hline 0.04 \\ 0.045 \\ \hline 0.02 \\ 0.045 \\ 0.02 \end{pmatrix}.$$

The system is exposed to random external shocks with PH representation $(\gamma, \mathbf{L})$ with $\gamma = (1,0)$ and $\mathbf{L} = \begin{pmatrix} -3 & 2.9 \\ 2.9 & -3 \end{pmatrix}$. The shock can provoke repairable or non-repairable failure according to these transition rates,

$$\mathbf{L}_r^0 = \begin{pmatrix} 0.08 \\ 0.08 \end{pmatrix} \text{ and } \mathbf{L}_{nr}^0 = \begin{pmatrix} 0.02 \\ 0.02 \end{pmatrix}.$$

The corrective repair and preventive maintenance time are phase type distributed with representation $(\boldsymbol{\beta}^1, \mathbf{S}_1)$ and $(\boldsymbol{\beta}^2, \mathbf{S}_2)$ respectively, where

$$\boldsymbol{\beta}^1 = (1,0), \qquad \boldsymbol{\beta}^2 = (1,0),$$

$$\mathbf{S}_1 = \begin{pmatrix} -1 & 0.5 \\ 0.5 & -1 \end{pmatrix}, \quad \mathbf{S}_2 = \begin{pmatrix} -2 & 0.005 \\ 0.005 & -2 \end{pmatrix}.$$

Rewards and cost are introduced in the problem. A profit per unit of time equal to $B = 100$ monetary units (m.u.) occurs whereas the system is operational (equal cost when it is not operational, $A=100$). A cost is produced depending on the operational degradation level while it is operational; 0.1 m,u,, 0.5 m.u. and 1 m.u. respectively for each one. Each unit of time that the repairperson is idle at the workplace, a cost equal to $r_S = 0.5$ m.u. is produced. This amount increases when the repairperson is working. If the repairperson is engaged in preventive maintenance, the cost increases by 1.5 m.u. and 9.5 m.u. for corrective repair.

In the following, we examine how the repairperson's vacation time should be distributed to optimise net rewards. To do so, it is assumed that the distribution of the vacation time is phase type (gamma distribution) with representation,

$$\mathbf{v} = (1,0); \ \mathbf{V} = \begin{pmatrix} -\lambda_1 & \lambda_1 \\ 0 & -\lambda_2 \end{pmatrix}.$$

To get the optimum model, the net reward profit in stationary regime is maximised,

$$\hat{\lambda}_1, \hat{\lambda}_2 \text{ s.t. } \Gamma(\hat{\lambda}_1, \hat{\lambda}_2) = \sup_{\lambda_1, \lambda_2} \Gamma(\lambda_1, \lambda_2).$$

These values are $\hat{\lambda}_1 = 5.8003$ and $\hat{\lambda}_2 = 5.8003$ for the case with preventive maintenance (maximum net profit in stationary regime 2.0943 m.u per unit of time) and $\hat{\lambda}_{1,smp} = 5.4502$ and $\hat{\lambda}_{2,smp} = 5.4502$ for the scenario without preventive maintenance.

For both optimum systems, the cumulative net profit per unit of time is compared in Figure 1. It is observed that the system with preventive maintenance is always better than the system without preventive maintenance from an economic point of view. The system without preventive maintenance is in deficit at any time, but when preventive maintenance is included in the system it is profitable from time 433.17.

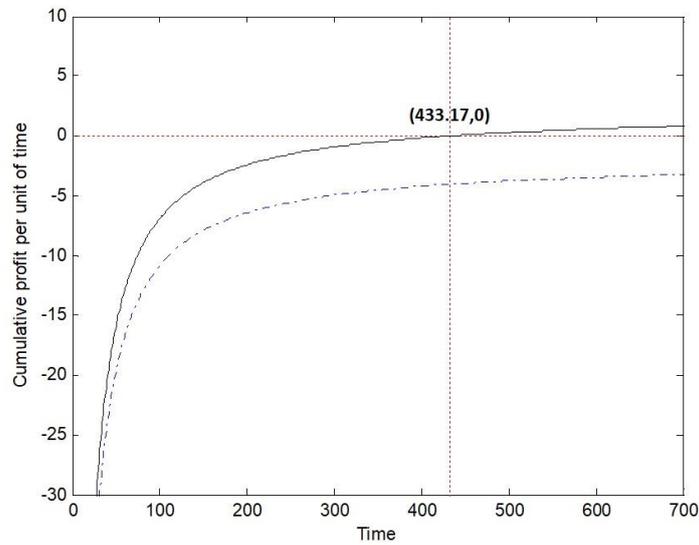

Figure 1. Cumulative net profit per unit of time over time (with preventive maintenance, continuous line; without preventive maintenance, dashed line)

Several performance measures such as the availability and mean events time have been worked out. Figure 1 shows the availability for both cases.

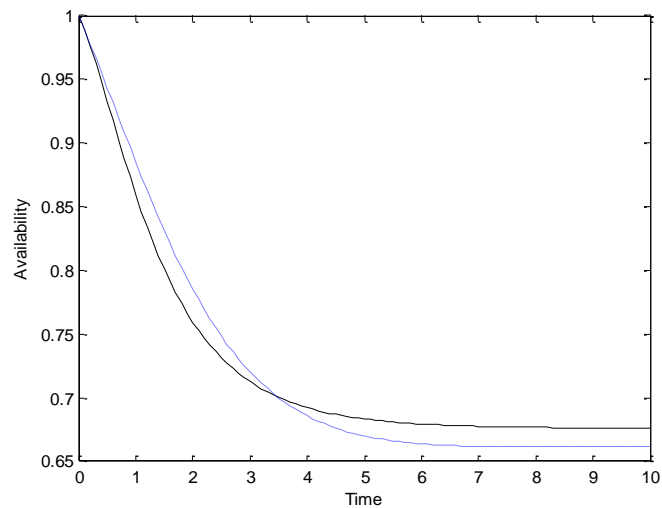

Figure 2. Availability for the optimum systems (with preventive maintenance, continuous line; without preventive maintenance, dashed line)

Table 1 shows the stationary distribution for both models. These values can be interpreted as the proportion of time in each macro-state.

| $\boldsymbol{\pi}_{O_1} \cdot \mathbf{e}$ | $\boldsymbol{\pi}_{O_2^{WR}} \cdot \mathbf{e}$ | $\boldsymbol{\pi}_{O_2^{R}} \cdot \mathbf{e}$ | $\boldsymbol{\pi}_{O_3^{WR}} \cdot \mathbf{e}$ | $\boldsymbol{\pi}_{RF} \cdot \mathbf{e}$ | $\boldsymbol{\pi}_{NRF} \cdot \mathbf{e}$ | $\boldsymbol{\pi}_{PM} \cdot \mathbf{e}$ | $\boldsymbol{\pi}_{CR} \cdot \mathbf{e}$ |
|---|---|---|---|---|---|---|---|
| 0.3851 | 0.0502 | 0.2387 | 0.0038 | 0.0133 | 0.0030 | 0.0479 | 0.2581 |
| (0.2909) | (0.0407) | (0.3304) | | (0.0100) | (0.0023) | | (0.3257) |

Table 1. Stationary distribution by considering the macro-states (without preventive maintenance in parenthesis)

Finally, several measures as ROCOF and mean number of events in transient and stationary regime are compared in Table 2.

| | $t=1$ | $t=5$ | $t=10$ | $t=50$ | | $t=\infty$ |
|---|---|---|---|---|---|---|
| $ROCOF_{RF}(t)$ | 0.1423 | 0.1315 | 0.1291 | 0.1290 | $ROCOF_{RF}$ | 0.1290 |
| | (0.1602) | (0.1688) | (0.1628) | (0.1629) | | (0.1629) |
| $ROCOF_{NRF}(t)$ | 0.0292 | 0.0263 | 0.0259 | 0.0259 | $ROCOF_{NRF}$ | 0.0259 |
| | (0.0308) | (0.0272) | (0.0263) | (0.0264) | | (0.0264) |
| $MN^{RF}(t)$ | 0.1201 | 0.6764 | 1.3247 | 6.4860 | $MN^{RF}$ | 0.1290 |
| | (0.1262) | (0.8332) | (1.6540) | (8.1686) | | (0.1629) |
| $MN^{NRF}(t)$ | 0.0261 | 0.1376 | 0.2676 | 1.3027 | $MN^{NRF}$ | 0.0259 |
| | (0.0266) | (0.1458) | (0.2782) | (1.3326) | | (0.0264) |
| $MN^{PM}(t)$ | 0.0487 | 0.4614 | 0.9429 | 4.7694 | $MN^{PM}$ | 0.0957 |
| $MN^{CR}(t)$ | 0.0978 | 0.6631 | 1.3114 | 6.4727 | $MN^{CR}$ | 0.1290 |
| | (0.1042) | (0.8235) | (1.6440) | (8.1586) | | (0.1629) |
| $MN^{I}(t)$ | 2.1841 | 7.6818 | 13.8847 | 63.5153 | $MN^{I}$ | 1.2408 |
| | (2.1750) | (6.6759) | (11.3160) | (48.7966) | | (0.9372) |
| $MN^{NU}(t)$ | 0.0210 | 0.1347 | 0.2646 | 1.2997 | $MN^{NU}$ | 0.0210 |
| | (0.0217) | (0.1436) | (0.2760) | (1.3303) | | (0.0264) |

Table 2. ROCOF and mean number of events (without preventive maintenance in parenthesis)

# 7. Conclusions

This paper presents two complex multi-state systems subject to various types of failure where in one of them preventive maintenance is applied. These systems are composed of several internal degradation levels and are subject to internal failure and random external shocks. The possible internal failure and/or external shocks may provoke repairable or non-repairable failure. A vacation policy is included in the model to optimize it, considering a financial point of view. Both systems are modelled using a Markovian Arrival Process with Marked arrivals in an algorithmic and computational form.

It is shown that the PH and MMAP enable us to express the modelling and its associated measures in a well structured way. Costs and rewards are included in the model and several associated measures are worked out. One interesting measure, the net reward per unit of time function, is built and it is considered to optimize systems according to vacation time distribution. A numerical example, optimizing similar systems with and without preventive maintenance, and comparing them, illustrates the versatility of the model.

**Acknowledgements**

This paper is supported by the project FQM-307 of the Government of Andalusia (Spain), by the project PID2020-113961GB-I00 of the Spanish Ministry of Science and Innovation (also supported by the European Regional Development Fund program, ERDF) and the project A-FQM-66-UGR20 of the Ministry of Knowledge, Research and University, Junta de Andalucía (Spain). Also, the authors acknowledge financial support by the IMAG–María de Maeztu grant CEX2020-001105-M / AEI / 10.13039/501100011033.

**APPENDIX A**

In this appendix, the matrix blocks and the stationary regime is given for the case without preventive maintenance addressed in Section 2.1. The state space is described in that section. The events associated with this system are

**Events**

0: No events (no PM, no failure, no end of vacation)

*RF+CR*: Repairable failure and corrective repair

*RF*: Repairable failure without immediate corrective repair

*NRF+NU*: Non-repairable failure and new unit

*NRF*: Non-repairable failure without immediate new unit

*I*: return of a vacation period

*I+CR*: return of a vacation period and corrective repair

*I+NU*: return and new unit

Therefore, the MMAP has the following representation
$$\left(\boldsymbol{D}^0, \boldsymbol{D}^{RF+CR}, \boldsymbol{D}^{RF}, \boldsymbol{D}^{NRF+NU}, \boldsymbol{D}^{NRF}, \boldsymbol{D}^{I}, \boldsymbol{D}^{I+CR}, \boldsymbol{D}^{I+NU}\right).$$

The block matrices are

$$\mathbf{D}^O = \begin{pmatrix} \mathbf{C}^1_{O_1 O_1} & \mathbf{C}_{O_1 O_2^{WR}} & 0 & 0 & 0 & 0 \\ 0 & \mathbf{C}_{O_2^{WR} O_2^{WR}} & 0 & 0 & 0 & 0 \\ 0 & 0 & \mathbf{C}_{O_2^R O_2^R} & 0 & 0 & 0 \\ 0 & 0 & 0 & \mathbf{C}_{RF^{WR} RF^{WR}} & 0 & 0 \\ 0 & 0 & 0 & 0 & \mathbf{C}_{NRF^{WR} NRF^{WR}} & 0 \\ \mathbf{C}_{CRO_1} & 0 & 0 & 0 & 0 & \mathbf{C}_{CRCR} \end{pmatrix}$$

$$\mathbf{D}^{RF+CR} = \begin{pmatrix} 0 & 0 & 0 & 0 & 0 & 0 \\ 0 & 0 & 0 & 0 & 0 & 0 \\ 0 & 0 & 0 & 0 & 0 & \mathbf{C}_{O_2^R CR} \\ 0 & 0 & 0 & 0 & 0 & 0 \\ 0 & 0 & 0 & 0 & 0 & 0 \\ 0 & 0 & 0 & 0 & 0 & 0 \end{pmatrix}, \quad \mathbf{D}^{RF} = \begin{pmatrix} 0 & 0 & 0 & \mathbf{C}_{O_1 RF^{WR}} & 0 & 0 \\ 0 & 0 & 0 & \mathbf{C}_{O_2^{WR} RF^{WR}} & 0 & 0 \\ 0 & 0 & 0 & 0 & 0 & 0 \\ 0 & 0 & 0 & 0 & 0 & 0 \\ 0 & 0 & 0 & 0 & 0 & 0 \\ 0 & 0 & 0 & 0 & 0 & 0 \end{pmatrix}$$

$$\mathbf{D}^{NRF+NU} = \begin{pmatrix} 0 & 0 & 0 & 0 & 0 & 0 \\ 0 & 0 & 0 & 0 & 0 & 0 \\ \mathbf{C}_{O_2^R O_1} & 0 & 0 & 0 & 0 & 0 \\ 0 & 0 & 0 & 0 & 0 & 0 \\ 0 & 0 & 0 & 0 & 0 & 0 \\ 0 & 0 & 0 & 0 & 0 & 0 \end{pmatrix} \quad \mathbf{D}^{NRF} = \begin{pmatrix} 0 & 0 & 0 & 0 & \mathbf{C}_{O_1 NRF^{WR}} & 0 \\ 0 & 0 & 0 & 0 & \mathbf{C}_{O_2^{WR} NRF^{WR}} & 0 \\ 0 & 0 & 0 & 0 & 0 & 0 \\ 0 & 0 & 0 & 0 & 0 & 0 \\ 0 & 0 & 0 & 0 & 0 & 0 \\ 0 & 0 & 0 & 0 & 0 & 0 \end{pmatrix}$$

$$\mathbf{D}^{I} = \begin{pmatrix} \mathbf{C}^2_{O_1 O_1} & 0 & 0 & 0 & 0 & 0 \\ 0 & 0 & \mathbf{C}_{O_2^{WR} O_2^{R}} & 0 & 0 & 0 \\ 0 & 0 & 0 & 0 & 0 & 0 \\ 0 & 0 & 0 & 0 & 0 & 0 \\ 0 & 0 & 0 & 0 & 0 & 0 \\ 0 & 0 & 0 & 0 & 0 & 0 \end{pmatrix}, \mathbf{D}^{I+CR} = \begin{pmatrix} 0 & 0 & 0 & 0 & 0 & 0 \\ 0 & 0 & 0 & 0 & 0 & 0 \\ 0 & 0 & 0 & 0 & 0 & 0 \\ 0 & 0 & 0 & 0 & 0 & \mathbf{C}_{RF^{WR} CR} \\ 0 & 0 & 0 & 0 & 0 & 0 \\ 0 & 0 & 0 & 0 & 0 & 0 \end{pmatrix},$$

$$\mathbf{D}^{I+NU} = \begin{pmatrix} 0 & 0 & 0 & 0 & 0 & 0 \\ 0 & 0 & 0 & 0 & 0 & 0 \\ 0 & 0 & 0 & 0 & 0 & 0 \\ 0 & 0 & 0 & 0 & 0 & 0 \\ \mathbf{C}_{NRFO_1} & 0 & 0 & 0 & 0 & 0 \\ 0 & 0 & 0 & 0 & 0 & 0 \end{pmatrix},$$

where the matrix-blocks are

$\mathbf{C}^1_{O_1 O_1} = \mathbf{T}_{11} \oplus \mathbf{L} \otimes \mathbf{I} + \mathbf{I} \otimes \mathbf{I} \otimes \mathbf{V}$ ; $\mathbf{C}^2_{O_1 O_1} = \mathbf{I} \otimes \mathbf{I} \otimes \mathbf{V}^0 \boldsymbol{\upsilon}$ ; $\mathbf{C}_{O_1 O_2^{WR}} = \mathbf{T}_{12} \otimes \mathbf{I} \otimes \mathbf{I}$ ;

$\mathbf{C}_{O_1 RF^{WR}} = \mathbf{T}^0_{1,r} \otimes \mathbf{I} \otimes \mathbf{I} + \mathbf{e} \otimes \mathbf{L}^0_r \boldsymbol{\gamma} \otimes \mathbf{I}$ ; $\mathbf{C}_{O_1 NRF^{WR}} = \mathbf{T}^0_{1,nr} \otimes \mathbf{I} \otimes \mathbf{I} + \mathbf{I} \otimes \mathbf{L}^0_{nr} \boldsymbol{\gamma} \otimes \mathbf{I}$

$\mathbf{C}_{O_2^{WR} O_2^{WR}} = \mathbf{T}_{22} \oplus \mathbf{L} \otimes \mathbf{I} + \mathbf{I} \otimes \mathbf{I} \otimes \mathbf{V}$ ; $\mathbf{C}_{O_2^{WR} O_2^{R}} = \mathbf{I} \otimes \mathbf{I} \otimes \mathbf{V}^0$ ;

$\mathbf{C}_{O_2^{WR} RF^{WR}} = \mathbf{T}^0_{2,r} \otimes \mathbf{I} \otimes \mathbf{I} + \mathbf{e} \otimes \mathbf{L}^0_r \boldsymbol{\gamma} \otimes \mathbf{I}$ ; $\mathbf{C}_{O_2^{WR} NRF^{WR}} = \mathbf{T}^0_{2,nr} \otimes \mathbf{I} \otimes \mathbf{I} + \mathbf{e} \otimes \mathbf{L}^0_{nr} \boldsymbol{\gamma} \otimes \mathbf{I}$ ;

$\mathbf{C}_{O_2^{R} O_1} = \mathbf{T}^0_{2,nr} \otimes \boldsymbol{\alpha}_1 \otimes \mathbf{I} \otimes \boldsymbol{\nu} + \mathbf{e} \otimes \boldsymbol{\alpha}_1 \otimes \mathbf{L}^0_{nr} \boldsymbol{\gamma} \otimes \boldsymbol{\upsilon}$ ; $\mathbf{C}_{O_2^{R} O_2^{R}} = \mathbf{T}_{22} \otimes \mathbf{I} + \mathbf{I} \otimes \mathbf{L}$ ;

$\mathbf{C}_{O_2^{R} CR} = \mathbf{T}^0_{2,r} \otimes \mathbf{I} \otimes \boldsymbol{\beta}^1 + \mathbf{e} \otimes \mathbf{L}^0_r \boldsymbol{\gamma} \otimes \boldsymbol{\beta}^1$ ; $\mathbf{C}_{RF^{WR} RF^{WR}} = (\mathbf{L} + \mathbf{L}^0 \boldsymbol{\gamma}) \oplus \mathbf{V}$ ; $\mathbf{C}_{RF^{WR} CR} = \mathbf{I} \otimes \mathbf{V}^0 \otimes \boldsymbol{\beta}^1$ ;

$\mathbf{C}_{NRF^{WR} O_1} = \boldsymbol{\alpha}_1 \otimes \mathbf{I} \otimes \mathbf{V}^0 \boldsymbol{\upsilon}$ ; $\mathbf{C}_{NRF^{WR} NRF^{WR}} = (\mathbf{L} + \mathbf{L}^0 \boldsymbol{\gamma}) \oplus \mathbf{V}$ ; $\mathbf{C}_{CRO_1} = \boldsymbol{\alpha}_1 \otimes \mathbf{I} \otimes \boldsymbol{\upsilon} \otimes \mathbf{S}^0_1$ ;

$\mathbf{C}_{CRCR} = (\mathbf{L} + \mathbf{L}^0 \boldsymbol{\gamma}) \oplus \mathbf{S}_1$ .

**Stationary distribution**

The stationary distribution for the case without preventive maintenance is also partitioned according to the macro-state space *S*, composed of six macro-states in this case. The process generator is denoted as

$$\mathbf{D} = \begin{pmatrix} D_{11} & D_{12} & 0 & D_{14} & D_{15} & 0 \\ 0 & D_{22} & D_{23} & D_{24} & D_{25} & 0 \\ D_{31} & 0 & D_{33} & 0 & 0 & D_{36} \\ 0 & 0 & 0 & D_{44} & 0 & D_{46} \\ D_{51} & 0 & 0 & 0 & D_{55} & 0 \\ D_{61} & 0 & 0 & 0 & 0 & D_{66} \end{pmatrix}.$$

The balance equations are given by

$$\pi_1 D_{11} + \pi_3 D_{31} + \pi_5 D_{51} + \pi_6 D_{61} = 0$$
$$\pi_1 D_{12} + \pi_2 D_{22} = 0$$
$$\pi_2 D_{23} + \pi_3 D_{33} = 0$$
$$\pi_1 D_{14} + \pi_2 D_{24} + \pi_4 D_{44} = 0$$
$$\pi_1 D_{15} + \pi_2 D_{25} + \pi_5 D_{55} = 0$$
$$\pi_3 D_{36} + \pi_4 D_{46} + \pi_6 D_{66} = 0$$
$$\boldsymbol{\pi} \cdot \mathbf{e} = 1$$

The solution of this matrix system is given by

$$\boldsymbol{\pi}_i = \boldsymbol{\pi}_1 \mathbf{R}_i \quad ; \quad i = 2,\ldots,6$$

with

$$\mathbf{R}_2 = \mathbf{G}_{12}, \ \mathbf{R}_3 = \mathbf{R}_2 \mathbf{G}_{23}, \ \mathbf{R}_4 = \mathbf{G}_{14} + \mathbf{R}_2 \mathbf{G}_{24}, \ \mathbf{R}_5 = \mathbf{G}_{15} + \mathbf{R}_2 \mathbf{G}_{25}, \ \mathbf{R}_6 = \mathbf{R}_3 \mathbf{G}_{36} + \mathbf{R}_4 \mathbf{G}_{46},$$

where $\mathbf{G}_{jk} = -\mathbf{D}_{jk} \mathbf{D}_{kk}^{-1}$ for the corresponding case.

The $\boldsymbol{\pi}_1$ vector is achieved from the first and last matrix equation,

$$\pi_1 D_{11} + \pi_3 D_{31} + \pi_5 D_{51} + \pi_6 D_{61} = 0$$
$$\boldsymbol{\pi} \cdot \mathbf{e} = 1$$

Then,

$$\boldsymbol{\pi}_1 = (0,1) \left[ \mathbf{A}^* \mid \left( \mathbf{I} + \sum_{a=2}^{6} \mathbf{R}_a \right) \mathbf{e} \right]^{-1}$$ with $\mathbf{A}^*$ being the matrix $\mathbf{A}$ without the first column and

$$\mathbf{A} = \mathbf{D}_{11} + \mathbf{R}_3 \mathbf{D}_{31} + \mathbf{R}_5 \mathbf{D}_{51} + \mathbf{R}_6 \mathbf{D}_{61}.$$

## APPENDIX B

This appendix contains the rest of the block-matrices for the system with preventive maintenance described in Section 3.1.

$$\mathbf{D}^O = \begin{pmatrix} \mathbf{C}^1_{O_1 O_1} & \mathbf{C}_{O_1 O_2^{WR}} & 0 & \mathbf{C}_{O_1 O_3^{WR}} & 0 & 0 & 0 & 0 \\ 0 & \mathbf{C}_{O_2^{WR} O_2^{WR}} & 0 & \mathbf{C}_{O_2^{WR} O_3^{WR}} & 0 & 0 & 0 & 0 \\ 0 & 0 & \mathbf{C}_{O_2^R O_2^R} & 0 & 0 & 0 & 0 & 0 \\ 0 & 0 & 0 & \mathbf{C}_{O_3^{WR} O_3^{WR}} & 0 & 0 & 0 & 0 \\ 0 & 0 & 0 & 0 & \mathbf{C}_{RF^{WR} RF^{WR}} & 0 & 0 & 0 \\ 0 & 0 & 0 & 0 & 0 & \mathbf{C}_{NRF^{WR} NRF^{WR}} & 0 & 0 \\ \mathbf{C}_{PMO_1} & 0 & 0 & 0 & 0 & 0 & \mathbf{C}_{PMPM} & 0 \\ \mathbf{C}_{CRO_1} & 0 & 0 & 0 & 0 & 0 & 0 & \mathbf{C}_{CRCR} \end{pmatrix}$$

$$\mathbf{D}^{PM} = \begin{pmatrix} 0 & 0 & 0 & 0 & 0 & 0 & 0 & 0 \\ 0 & 0 & 0 & 0 & 0 & 0 & 0 & 0 \\ 0 & 0 & 0 & 0 & 0 & 0 & \mathbf{C}_{O_2^R PM} & 0 \\ 0 & 0 & 0 & 0 & 0 & 0 & 0 & 0 \\ 0 & 0 & 0 & 0 & 0 & 0 & 0 & 0 \\ 0 & 0 & 0 & 0 & 0 & 0 & 0 & 0 \\ 0 & 0 & 0 & 0 & 0 & 0 & 0 & 0 \\ 0 & 0 & 0 & 0 & 0 & 0 & 0 & 0 \end{pmatrix} \quad \mathbf{D}^{I+PM} = \begin{pmatrix} 0 & 0 & 0 & 0 & 0 & 0 & 0 & 0 \\ 0 & 0 & 0 & 0 & 0 & 0 & 0 & 0 \\ 0 & 0 & 0 & 0 & 0 & 0 & 0 & 0 \\ 0 & 0 & 0 & 0 & 0 & 0 & \mathbf{C}_{O_3^{WR} PM} & 0 \\ 0 & 0 & 0 & 0 & 0 & 0 & 0 & 0 \\ 0 & 0 & 0 & 0 & 0 & 0 & 0 & 0 \\ 0 & 0 & 0 & 0 & 0 & 0 & 0 & 0 \\ 0 & 0 & 0 & 0 & 0 & 0 & 0 & 0 \end{pmatrix};$$

$$\mathbf{D}^{I+NU} = \begin{pmatrix} 0 & 0 & 0 & 0 & 0 & 0 & 0 & 0 \\ 0 & 0 & 0 & 0 & 0 & 0 & 0 & 0 \\ 0 & 0 & 0 & 0 & 0 & 0 & 0 & 0 \\ 0 & 0 & 0 & 0 & 0 & 0 & 0 & 0 \\ 0 & 0 & 0 & 0 & 0 & 0 & 0 & 0 \\ \mathbf{C}_{NRF^{WR} O_1} & 0 & 0 & 0 & 0 & 0 & 0 & 0 \\ 0 & 0 & 0 & 0 & 0 & 0 & 0 & 0 \\ 0 & 0 & 0 & 0 & 0 & 0 & 0 & 0 \end{pmatrix} \quad \mathbf{D}^{NRF} = \begin{pmatrix} 0 & 0 & 0 & 0 & 0 & \mathbf{C}_{O_1 NRF^{WR}} & 0 & 0 \\ 0 & 0 & 0 & 0 & 0 & \mathbf{C}_{O_2^{WR} NRF^{WR}} & 0 & 0 \\ 0 & 0 & 0 & 0 & 0 & 0 & 0 & 0 \\ 0 & 0 & 0 & 0 & 0 & \mathbf{C}_{O_3^{WR} NRF^{WR}} & 0 & 0 \\ 0 & 0 & 0 & 0 & 0 & 0 & 0 & 0 \\ 0 & 0 & 0 & 0 & 0 & 0 & 0 & 0 \\ 0 & 0 & 0 & 0 & 0 & 0 & 0 & 0 \\ 0 & 0 & 0 & 0 & 0 & 0 & 0 & 0 \end{pmatrix}$$

$$\mathbf{D}^{NRF+NU} = \begin{pmatrix} 0 & 0 & 0 & 0 & 0 & 0 & 0 & 0 \\ 0 & 0 & 0 & 0 & 0 & 0 & 0 & 0 \\ \mathbf{C}_{O_2^R O_1} & 0 & 0 & 0 & 0 & 0 & 0 & 0 \\ 0 & 0 & 0 & 0 & 0 & 0 & 0 & 0 \\ 0 & 0 & 0 & 0 & 0 & 0 & 0 & 0 \\ 0 & 0 & 0 & 0 & 0 & 0 & 0 & 0 \\ 0 & 0 & 0 & 0 & 0 & 0 & 0 & 0 \\ 0 & 0 & 0 & 0 & 0 & 0 & 0 & 0 \end{pmatrix}$$

$$\mathbf{D}^{I} = \begin{pmatrix} \mathbf{C}_{O_1 O_1}^2 & 0 & 0 & 0 & 0 & 0 & 0 & 0 \\ 0 & 0 & \mathbf{C}_{O_2^{WR} O_2^R} & 0 & 0 & 0 & 0 & 0 \\ 0 & 0 & 0 & 0 & 0 & 0 & 0 & 0 \\ 0 & 0 & 0 & 0 & 0 & 0 & 0 & 0 \\ 0 & 0 & 0 & 0 & 0 & 0 & 0 & 0 \\ 0 & 0 & 0 & 0 & 0 & 0 & 0 & 0 \\ 0 & 0 & 0 & 0 & 0 & 0 & 0 & 0 \\ 0 & 0 & 0 & 0 & 0 & 0 & 0 & 0 \end{pmatrix} \quad \mathbf{D}^{I+CR} = \begin{pmatrix} 0 & 0 & 0 & 0 & 0 & 0 & 0 & 0 \\ 0 & 0 & 0 & 0 & 0 & 0 & 0 & 0 \\ 0 & 0 & 0 & 0 & 0 & 0 & 0 & 0 \\ 0 & 0 & 0 & 0 & 0 & 0 & 0 & 0 \\ 0 & 0 & 0 & 0 & 0 & 0 & 0 & \mathbf{C}_{RF^{WR} CR} \\ 0 & 0 & 0 & 0 & 0 & 0 & 0 & 0 \\ 0 & 0 & 0 & 0 & 0 & 0 & 0 & 0 \\ 0 & 0 & 0 & 0 & 0 & 0 & 0 & 0 \end{pmatrix}$$

.
.

$\mathbf{C}_{O_1 O_1}^1 = \mathbf{T}_{11} \oplus \mathbf{L} \otimes \mathbf{I} + \mathbf{I} \otimes \mathbf{I} \otimes \mathbf{V}$ ; $\mathbf{C}_{O_1 O_1}^2 = \mathbf{I} \otimes \mathbf{I} \otimes \mathbf{V}^0 \boldsymbol{\upsilon}$ ; $\mathbf{C}_{O_1 O_2^{WR}} = \mathbf{T}_{12} \otimes \mathbf{I} \otimes \mathbf{I}$ ;

$\mathbf{C}_{O_1 O_3^{WR}} = \mathbf{T}_{13} \otimes \mathbf{I} \otimes \mathbf{I}$ ; $\mathbf{C}_{O_1 NRF^{WR}} = \mathbf{T}_{1,nr}^0 \otimes \mathbf{I} \otimes \mathbf{I} + \mathbf{e} \otimes \mathbf{L}_{nr}^0 \boldsymbol{\gamma} \otimes \mathbf{I}$

$\mathbf{C}_{O_2^{WR} O_2^{WR}} = \mathbf{T}_{22} \oplus \mathbf{L} \otimes \mathbf{I} + \mathbf{I} \otimes \mathbf{I} \otimes \mathbf{V}$ ; $\mathbf{C}_{O_2^{WR} O_2^R} = \mathbf{I} \otimes \mathbf{I} \otimes \mathbf{V}^0$ ; $\mathbf{C}_{O_2^{WR} O_3^{WR}} = \mathbf{T}_{23} \otimes \mathbf{I} \otimes \mathbf{I}$ ;

$\mathbf{C}_{O_2^{WR} NRF^{WR}} = \mathbf{T}_{2,nr}^0 \otimes \mathbf{I} \otimes \mathbf{I} + \mathbf{e} \oplus \mathbf{L}_{nr}^0 \boldsymbol{\gamma} \otimes \mathbf{I}$ ; $\mathbf{C}_{O_2^R O_1} = \mathbf{T}_{2,nr}^0 \otimes \boldsymbol{\alpha}_1 \otimes \mathbf{I} \otimes \mathbf{v} + \mathbf{e} \otimes \boldsymbol{\alpha}_1 \otimes \mathbf{L}_{nr}^0 \boldsymbol{\gamma} \otimes \boldsymbol{\upsilon}$ ;

$\mathbf{C}_{O_2^R O_2^R} = \mathbf{T}_{22} \otimes \mathbf{I} + \mathbf{I} \otimes \mathbf{L}$ ; $\mathbf{C}_{O_2^R PM} = \mathbf{T}_{23} \mathbf{e} \otimes \mathbf{I} \otimes \boldsymbol{\beta}^2$ ; $\mathbf{C}_{O_3^{WR} O_3^{WR}} = \mathbf{T}_{33} \oplus \mathbf{L} \otimes \mathbf{I} + \mathbf{I} \otimes \mathbf{I} \otimes \mathbf{V}$ ;

$\mathbf{C}_{O_3^{WR} RF^{WR}} = \mathbf{T}_{3,r}^0 \otimes \mathbf{I} \otimes \mathbf{I} + \mathbf{e} \otimes \mathbf{L}_r^0 \boldsymbol{\gamma} \otimes \mathbf{I}$  $\mathbf{C}_{O_3^{WR} NRF^{WR}} = \mathbf{T}_{3,nr}^0 \otimes \mathbf{I} \otimes \mathbf{I} + \mathbf{e} \otimes \mathbf{L}_{nr}^0 \boldsymbol{\gamma} \otimes \mathbf{I}$ ;

$\mathbf{C}_{O_3^{WR} PM} = \mathbf{e} \otimes \mathbf{I} \otimes \mathbf{V}^0 \otimes \boldsymbol{\beta}^2$ ; $\mathbf{C}_{RF^{WR} RF^{WR}} = \left(\mathbf{L} + \mathbf{L}^0 \boldsymbol{\gamma}\right) \oplus \mathbf{V}$ ; $\mathbf{C}_{RF^{WR} CR} = \mathbf{I} \otimes \mathbf{V}^0 \otimes \boldsymbol{\beta}^1$ ;

$\mathbf{C}_{NRF^{WR} O_1} = \boldsymbol{\alpha}_1 \otimes \mathbf{I} \otimes \mathbf{V}^0 \boldsymbol{\upsilon}$ ; $\mathbf{C}_{NRF^{WR} NRF^{WR}} = \left(\mathbf{L} + \mathbf{L}^0 \boldsymbol{\gamma}\right) \oplus \mathbf{V}$ ; $\mathbf{C}_{PM O_1} = \boldsymbol{\alpha}_1 \otimes \mathbf{I} \otimes \boldsymbol{\upsilon} \otimes \mathbf{S}_2^0$ ;

$\mathbf{C}_{PMPM} = \left(\mathbf{L} + \mathbf{L}^0 \boldsymbol{\gamma}\right) \oplus \mathbf{S}_2$ ; $\mathbf{C}_{CR O_1} = \boldsymbol{\alpha}_1 \otimes \mathbf{I} \otimes \boldsymbol{\upsilon} \otimes \mathbf{S}_1^0$ ; $\mathbf{C}_{CRCR} = \left(\mathbf{L} + \mathbf{L}^0 \boldsymbol{\gamma}\right) \oplus \mathbf{S}_1$ .